\documentclass{epl}

\title{Multi-Agent Complex Systems and Many-Body Physics}
\author{Neil F. Johnson\inst{1} \and David M.D. Smith\inst{2} \and Pak Ming Hui\inst{3}}

\institute{                    
  \inst{1} Physics Department, Oxford University, Oxford, OX1 3PU, UK\\
  \inst{2} Mathematics Department, Oxford University, Oxford, OX1 2EL, UK\\
    \inst{3} Department of Physics, Chinese University of Hong Kong, Hong Kong
}

\begin{document}

\maketitle

\begin{abstract}
Multi-agent complex systems comprising populations of
decision-making particles, have many potential applications across the biological,
informational and social sciences. We show that the time-averaged dynamics in such systems bear a striking resemblance to conventional
many-body physics. For the specific example of the
Minority Game, this analogy enables us to obtain analytic expressions which are in excellent
agreement with numerical simulations. 
\end{abstract}

Multi-agent simulations are currently being used to study the
dynamical behavior within a wide variety of Complex Systems
\cite{casti}. Within these simulations, $N$ decision-making
particles or {\em agents} (e.g. commuters, traders, computer
programs \cite{econo,rob,book})
repeatedly compete with each other for some limited global
resource (e.g. road space, best buy/sell price, processing time) using sets of rules
which may differ between agents and may change in time. The
population is therefore competitive, heterogeneous and adaptive. A
simple version of such a scenario, which has generated more than one hundred
papers since 1998, is the Minority Game (MG) of Challet and Zhang
\cite{econo,rob,book,others,savit,emg,us,paul}. 

Here we show that the time-averaged dynamics of such multi-agent systems -- in particular, their $n$-point correlation functions -- can be interpreted as a generalization of conventional many-body physics \cite{mahan}. We also show that these correlation functions can be evaluated accurately if one regroups the agents into clusters of like particles (i.e. crowds) and their anti-correlated mirror-images (i.e. anticrowds).
When applied to the MG, this approach yields a set of analytic
results which are in excellent agreement with the numerical
findings of Savit {\em et al.} (see Fig. 1) \cite{savit}. Although there have been many other MG theories proposed to date
\cite{others}, none of these provides an analytic
description of this Savit-curve \cite{savit} (Fig. 1) over the entire
parameter space.

\begin{figure}
\onefigure[width=.5\textwidth]{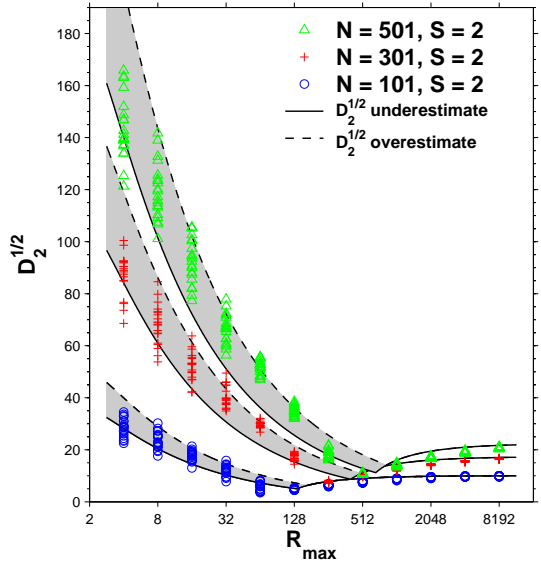}
\caption{Results for the standard deviation of fluctuations
$D_{2}^{1/2}$ in the Minority Game. Numerical results correspond to 20
different runs at each $N$ and $R_{\rm max}$. The theoretical curves
are generated using the analytic expressions in the text. The shaded area
bounded by the upper and lower curves shows our theoretical prediction of the
numerical spread for a given $N$.  In line with the original numerical results of
Ref.
\cite{savit}, we have chosen successive $R_{\rm max}$ tick-values to
increase by a factor of 4. }
\label{figure1}
\end{figure}

\begin{figure}
\onefigure[width=0.7\textwidth]{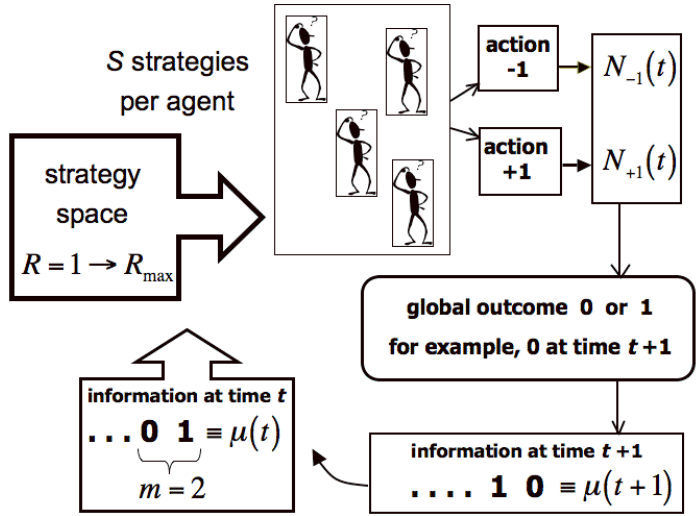}
\caption{General setup: At
timestep
$t$, each agent decides between action $-1$ and action $+1$ based on the
predictions of the $S$ strategies that he possesses. $N_{-1}[t]$ agents
choose
 $-1$, and $N_{+1}[t]$ choose $+1$. A global outcome 0 or 1 is assigned depending on the rules of the game (e.g. minority group wins). Strategies are
rewarded/penalized one virtual point according to whether their predicted action
would have been a winning/losing action.}
\label{figure2}
\end{figure}

\begin{figure}
\onefigure[width=0.7\textwidth]{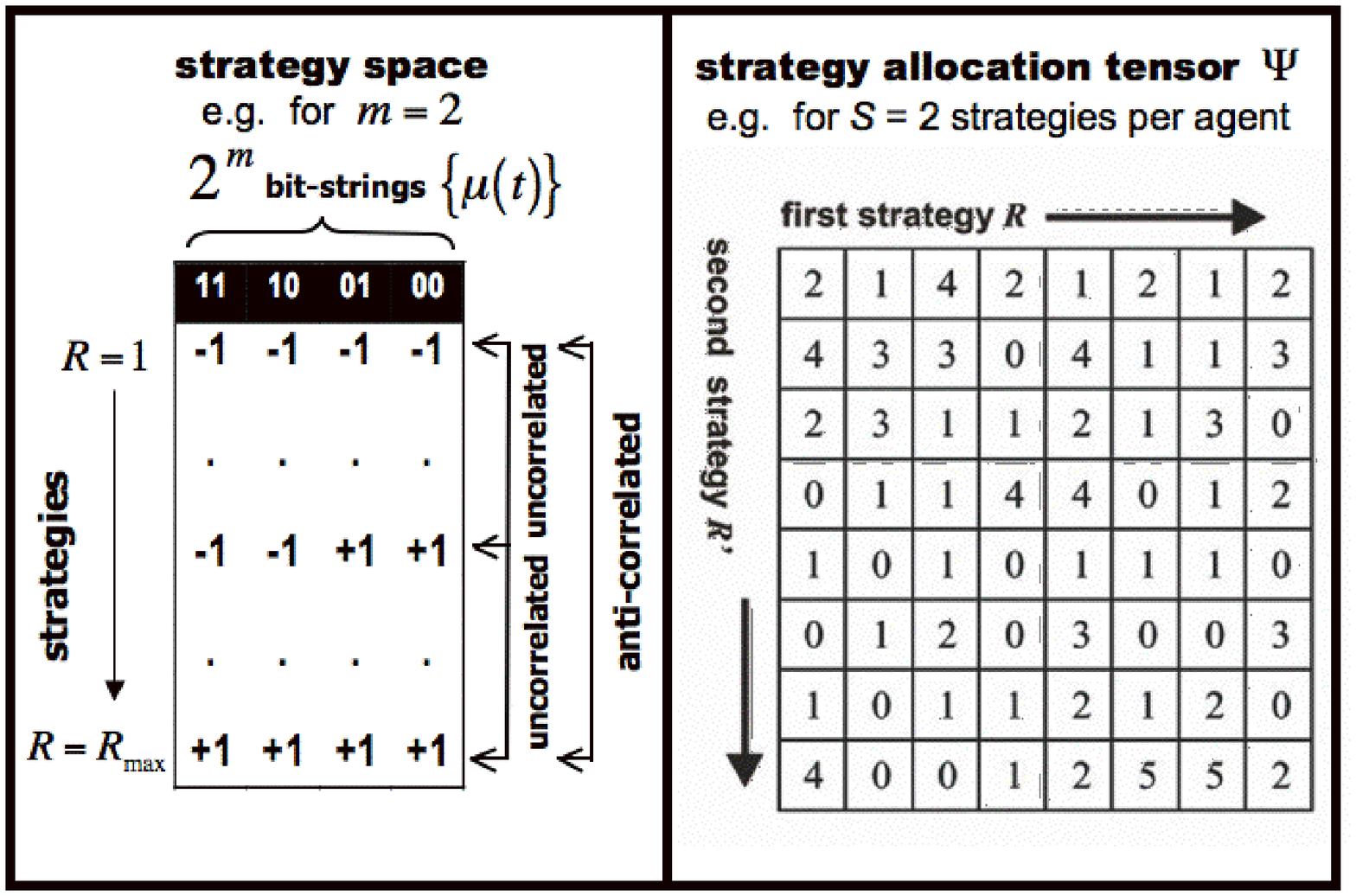}
\caption{Left: strategy space for $m=2$, together with some example strategies.
The reduced strategy space contains $2.{2^{m}}=2P$ strategies with specific correlations (i.e. fully correlated, uncorrelated or anti-correlated \cite{book,others}).
Right: strategy allocation tensor $\Psi$ in the case of $S=2$ strategies per agent. Each square shows the number of agents who were assigned strategy $R$ and then $R'$.}
\label{figure3}
\end{figure}

Figure 2 shows a generic setup of such a multi-agent system. The Minority Game represents a specific example of this setup, in which the identity of the minority group provides the global outcome -- however, what follows does not depend specifically on such a minority rule. There are $N$ agents (e.g. commuters) who repeatedly decide between
two actions at each timestep $t$ (e.g. $+1/-1 \equiv$ take route
A/B) using their individual $S$ strategies. The agents have access to some common information $\mu(t)$. For example, Fig. 2 shows $\mu(t)$ to be the past few outcomes of the system where 0 or 1 could represent route A  or B being the least crowded, respectively. Hence $\mu(t)$ is, in the setup of Fig. 2, a binary string of outcomes.
Each strategy, labelled $R$, comprises a particular action $a_{R}^{\mu (t)}=\pm 1$ for
each $\mu(t)\in \{\mu(t)\}$, and the set of all possible strategies
constitutes a strategy space $\Theta\equiv \{R\}$.  Figure 3 (left) gives an example of such a strategy space in the case that each agent has information $\mu(t)$ corresponding to the most recent 
$m=2$ outcomes.  Each agent has some subset  
$S$ of these strategies, hence the strategy
allocation among agents can be described in terms of a rank-$S$
tensor $\Psi$ \cite{paul} where each entry gives the number of
agents holding a particular combination of $S$ strategies. Figure 3 (right) gives an example of this strategy allocation matrix $\Psi$ for $S=2$.  We
assume $\Psi$ to be constant over the timescale for which
time-averages are taken. A single $\Psi$ `macrostate' corresponds
to many possible `microstates' describing the specific partitions
of strategies among the $N$ agents \cite{paul}. To allow for large
strategy spaces and large sets of global information, we can convert the strategies $\{R\}$ to their decimal equivalents and label them from
$R=1\rightarrow R_{\rm max}$, where $R_{\rm max}=2.2^m$, in order of increasing magnitude. Hence for the example strategies in Fig. 3, we can relabel strategy $[-1\ -1\ -1\ -1]$ as $R=1$, $[-1\ -1\ +1\ +1]$ as $R=4$ and $[+1\ +1\ +1\ +1]$ as $R=R_{\rm max}=8$. Hence $R$ mimics a `coordinate' in that if an agent is using a particular strategy $R$, the agent can be thought of as sitting on a line at coordinate $R$. For large strategy spaces, there will be many possible $R$-values -- for example, $m\geq 3$ means that there are $2.{2^{m}}\geq 16$ strategies in the reduced strategy space of Fig. 3.  The line of ordered strategies, and hence allowed $R$-values, will therefore appear quite dense. Given this, we will take the liberty of writing down sums over $R$ as integrals over $R$ in the rest of this paper.

We denote the number of agents
choosing $-1$ (e.g. take route A, or buy a given stock) and $+1$ (e.g. take route B, or sell a given stock) as $N_{-1}(t)$  and $N_{+1}(t)$ respectively. Many macroscopic quantities of interest can be written in terms of $N_{-1}(t)$ and $N_{+1}(t)$. For example, the excess
traffic on route A as compared to route B, or the excess number of buyers over sellers of a given stock and hence its price-change at time $t$, can be written as $D(t)=N_{+1}(t)-N_{-1}(t)$. 
Hence we will focus here on $D(t)$, which is given by:
\begin{equation} D(t)
\equiv \int_{R=1}^{R_{\rm max}} dR\  a_{R}^{\mu (t)}
n_{R}^{{\underline{S}(t);\Psi}}, \label{defD}
\end{equation}
where 
$n_{R}^{{\underline{S}(t);\Psi}}$ is the number of agents using strategy
$R$ at time $t$, and 
$\underline{S}(t)$ is the current score-vector denoting the
past performance of each strategy \cite{paul}. In particular, the element  ${S_R}(t)$ represents the net number of successful predictions of strategy $R$ up to time $t$.  Therefore, the combination of
$\underline{S}(t)$, $\Psi$ and the game rule (e.g. use strategy
with best or second-best performance to date) defines $n_{R}^{{\underline{S}(t);\Psi}}$. The action $a_{R}^{\mu (t)}=\pm 1$ is determined
uniquely by ${\mu(t)}$.

In conventional many-body physics, we are typically  interested in the following
statistical properties of macroscopic measurable quantities such as $D(t)$: (i) the moments of the probability
distribution function (PDF) of $D(t)$ (e.g. mean, variance,
kurtosis) and (ii) the correlation functions that are products of
$D(t)$ at different times $t_1=t$, $t_2=t+\tau$,
$t_3=t+\tau'$ etc. (e.g. autocorrelation). Numerical multi-agent
simulations typically average over time $t$ and then over
configurations $\{\Psi\}$. A general expression to generate all
such functions, is therefore {\small\begin{eqnarray}
&&D_P^{(\tau,\tau',\tau'',\dots)} \equiv \left\langle\left\langle
D(t_1) D(t_2) \dots D(t_P)\right\rangle_t \right\rangle_\Psi
\label{gen}\\ & = & \left\langle\left\langle
\int\dots\int_{1}^{R_{\rm max}}dR_1\ dR_2 \dots dR_P \
a_{R_1}^{\mu(t_1)} a_{R_2}^{\mu(t_2)} \dots a_{R_P}^{\mu(t_P)}
n_{R_1}^{\underline{S}(t_1);\Psi}
n_{R_2}^{\underline{S}(t_2);\Psi} \dots
n_{R_P}^{\underline{S}(t_P);\Psi} \right\rangle _{t} \right\rangle
_{\Psi} \nonumber \\ & \equiv & \int\dots\int_{1}^{R_{\rm
max}}dR_1\ dR_2 \dots dR_P \left\langle\left\langle
V^{(P)}(R_1,R_2,\dots,R_P;t_1,t_2,\dots,t_P)
n_{R_1}^{\underline{S}(t_1);\Psi}
n_{R_2}^{\underline{S}(t_2);\Psi}  \dots
n_{R_P}^{\underline{S}(t_P);\Psi} \right\rangle _{t} \right\rangle
_{\Psi} \nonumber
\end{eqnarray}}

\noindent where
\begin{equation} V^{(P)}(R_1,R_2\dots,R_P;t_1,t_2\dots,t_P)\equiv
a_{R_1}^{\mu(t_1)} a_{R_2}^{\mu(t_2)}  \dots a_{R_P}^{\mu(t_P)} \label{defV}
\end{equation} resembles a time-dependent, non-translationally invariant, {\em
$p$-body} interaction potential in ${\bf R}\equiv
(R_1,R_2\dots,R_P)$-space, between $p$ charge-densities
$\{n_{R_i}^{\underline{S}(t_i);\Psi}\}$ of like-minded agents.
Note that each charge-density $n_{R_i}^{\underline{S}(t_i);\Psi}$
now possesses internal degrees of freedom determined by
$\underline{S}(t)$ and $\Psi$. Since
$\{n_{R_i}^{\underline{S}(t_i);\Psi}\}$ are determined by the
game's rules, Eq. (\ref{gen}) can be applied to {\em any}
multi-agent game, not just the MG. We focus here on moments of the PDF
of $D(t)$ where $\{t_{i}\} \equiv t$ and hence $\{\tau\}=0$.
Discussion of temporal correlation functions such as the
autocorrelation $D_{2}^{(\tau)}$ will be reported elsewhere. We
consider explicitly the variance $D_{2}$ to demonstrate the
approach, noting that higher-order moments such as $D_{4}$ (i.e.
kurtosis) which classify the non-Gaussianity of the PDF, can be
treated in a similar way. The potential $V^{(P)}$ is insensitive
to the configuration-average over $\{\Psi\}$, hence the mean is
given by \cite{note}:
\begin{equation} D_{1} = \int_{R=1}^{R_{\rm max}}dR \  \
\left\langle V^{(1)}(R;t)\
\left\langle n_{R}^{\underline{S}(t);\Psi} \right\rangle _{\Psi}\right\rangle
_{t}\ .
\end{equation} If the game's output is unbiased, the averages yield $D_1=0$.
This condition is not necessary -- one can simply subtract $D_1^2$ from the
right hand side of the expression for $D_2$ below -- however we will take
$D_{1}=0$ for clarity. The variance $D_2$ measures the fluctuations of $D(t)$
about its average value:
\begin{equation} D_2 = \int\int_{R,R^{\prime }=1}^{R_{\rm max}}dR dR^{\prime
}\  \left\langle V^{(2)}(R,R^{\prime};t)\
\left\langle n_{R}^{\underline{S}(t);\Psi} n_{R^{\prime
}}^{\underline{S}(t);\Psi}\right\rangle _{\Psi}\right\rangle _{t} \label{var}
\end{equation} where $V^{(2)}(R,R^{\prime};t)\equiv  a_{R}^{\mu (t)}
a_{R^{\prime }}^{\mu (t)}$ acts like a time-dependent, non-translationally
invariant, two-body interaction potential in
$(R,R')$-space.

These effective charge-densities and potential fluctuate in time.
It is reasonable to assume that the charge densities fluctuate around some mean
value, hence
$n_{R}^{\underline{S}(t);\Psi} n_{R'}^{\underline{S}(t);\Psi}=n_{R}
n_{R'}+\varepsilon_{R R'}^{\underline{S}(t);\Psi}(t)$ with mean $n_R n_{R'}$
plus a fluctuating term $\varepsilon_{R R'}^{\underline{S}(t);\Psi}(t)$. This
is a good approximation if we take $R$ to be a popularity-ranking (i.e. the
$R$th most popular strategy) or a strategy-performance ranking (i.e. the $R$th
best-performing strategy) since in these cases $n_{R}^{\underline{S}(t);\Psi}$
will be reasonably constant \cite{book}. For example, taking $R$ as a popularity-ranking
implies $n_{R=1}^{\underline{S}(t);\Psi}\geq n_{R=2}^{\underline{S}(t);\Psi}\geq
n_{R=3}^{\underline{S}(t);\Psi}\geq \dots$, thereby constraining the magnitude
of the fluctuations in the charge-density
$n_{R}^{\underline{S}(t);\Psi}$.  Hence
\begin{equation} D_2 = \int\int_{R,R^{\prime }=1}^{R_{\rm max}}dR dR^{\prime
}\  \left\langle V^{(2)}(R,R^{\prime};t)\ \left\langle n_{R}
n_{R'}+\varepsilon_{R R'}^{\underline{S}(t);\Psi}(t)
\right\rangle _{\Psi} \right\rangle _{t}\ .
\end{equation} We will assume that
$\varepsilon_{RR'}^{\underline{S}(t);\Psi}(t)$ averages out to
zero. In the presence of network connections between agents, there
can be strong correlations between these noise terms
$\varepsilon_{R R'}^{\underline{S}(t);\Psi}(t)$ and the
time-dependence of $V^{(2)}(R,R^{\prime};t)$, implying that the
averaging over $t$ should be carried out timestep-by-timestep \cite{lo}. For MG-like games without connections, the agents
cannot suddenly access larger numbers of strategies and hence
these correlations can be ignored. This gives
\begin{equation} D_2 =
\int\int_{R,R^{\prime }=1}^{R_{\rm max}}dR dR^{\prime }\  \left\langle
V^{(2)}(R,R^{\prime};t) \right\rangle _{t} \ n_{R} n_{R'}\ .\label{D2mid}
\end{equation} As in conventional many-body theory, the expectation value in Eq.
(\ref{D2mid}) can be `contracted' down by making use of the
equal-time correlations between $\{a^{\mu(t)}_{R}\}$. As is well-known
for MG-like games \cite{econo,others,us}, we can safely work in the so-called reduced strategy space which is constructed such that any
pair $R$ and $R'$  are either (i) fully correlated, i.e.
$a^{\mu(t)}_R=a^{\mu(t)}_{R'}$ for all  $\mu(t)$;
(ii) anti-correlated, i.e. $a^{\mu(t)}_R=-a^{\mu(t)}_{R'}$ for all
 $\mu(t)$;  (iii) uncorrelated, i.e.
$a^{\mu(t)}_R=a^{\mu(t)}_{R'}$ for half of
$\{\mu(t)\}$ while $a^{\mu(t)}_R=-a^{\mu(t)}_{R'}$ for the other
half of $\{\mu(t)\}$. In other words, one can choose two subsets of the strategy space
$\Theta$, i.e. $\Theta=U\oplus \overline U$, such that the
strategies within $U$ are uncorrelated, the strategies within
$\overline U$ are uncorrelated, the anticorrelated strategy of
$R\in U$ appears in $\overline U$, and the anticorrelated strategy
of $R\in \overline U$ appears in $U$. We can therefore break up
the integrals in Eq. (\ref{D2mid}): (i) $R'\equiv
R$ (i.e. fully-correlated) hence $\frac{1}{\mu_{\rm
max}}\int_{\mu=1}^{\mu_{\rm max}} d\mu a_{R}^{\mu} a_{R'}^{\mu}
=1$ and $\left\langle V^{(2)}(R,R^{\prime};t)\right\rangle
_{t}=1$; (ii) $R'\equiv {\overline R}$ (i.e. anticorrelated) which
yields $\frac{1}{\mu_{\rm max}}\int_{\mu=1}^{\mu_{\rm max}} d\mu
a_{R}^{\mu} a_{ R'}^{\mu} =-1$. If all possible global information
values $\{\mu\}$ are visited reasonably equally over a long
time-period, this implies $\left\langle
V^{(2)}(R,R^{\prime};t)\right\rangle _{t}=-1$. For the MG, for
example, $\{\mu\}$ corresponds to the $m$-bit histories which
indeed are visited equally for small $m$. For large $m$, they are
not visited equally for a given $\Psi$, but are when averaged over
all $\Psi$. If, by contrast, we happened to be considering some
general non-MG game where the $\mu$'s occur with unequal
probabilities $\rho_\mu$, even after averaging over all $\Psi$,
one can simply redefine the strategy subsets $U$ and $\overline U$
to yield a generalized scalar product, i.e. $\frac{1}{\mu_{\rm
max}}\int_{\mu=1}^{\mu_{\rm max}} d\mu a_{R}^{\mu} a_{R'}^{\mu} \
\rho_\mu=-1$ (or $0$ in case (iii)). (iii) $R'\perp R$ (i.e.
uncorrelated) which yields $\frac{1}{\mu_{\rm
max}}\int_{\mu=1}^{\mu_{\rm max}} d\mu a_{R}^{\mu} a_{R'}^{\mu}
=0$ and hence $\left\langle V^{(2)}(R,R^{\prime};t)\right\rangle
_{t}=0$. Hence
\begin{eqnarray} D_2 & = & \int\int_{R,R^{\prime }=1}^{R_{\rm max}}dR
dR^{\prime }\  \left\langle V^{(2)}(R,R^{\prime};t)\right\rangle _{t}\ n_{R}
n_{R'}=
\int_{R=1}^{R_{\rm max}}dR \  (n_{R} n_{R}-n_{R} n_{\overline R})\nonumber \\ &
= &
\int_{R\in U} dR \  (n_{R} n_{R}-n_{R} n_{\overline R}+ n_{\overline R}
n_{\overline R}-n_{\overline R} n_{R})
\nonumber \\ &
 = &
\int_{R\in U} dR \  (n_{R} - n_{\overline R})^2 \ .
\label{final}
\end{eqnarray} Equation (\ref{final}) must be evaluated together with the
condition which guarantees that the total number of agents $N$ is conserved:
\begin{equation} N=\int_{R=1}^{R_{\rm max}}dR \  n_{R}\equiv \int_{R\in U} dR
\  (n_{R} + n_{\overline R}) \ .
\label{vol}
\end{equation} Equation (\ref{final}) has a simple interpretation. Since
$n_{R}$ and $n_{\overline R}$ have opposite sign, they act like two
charge-densities of opposite charge which tend to cancel each other out: $n_{R}$ represents
a Crowd of like-minded people, while $n_{\overline R}$
corresponds to a like-minded Anticrowd who always do exactly the {\em opposite} of the
Crowd regardless of the specific $\mu(t)$. We have
effectively renormalized the charge-densities
$n_{R}^{\underline{S}(t);\Psi}$ and
$n_{R'}^{\underline{S}(t);\Psi}$ and their time- and position-dependent
two-body interaction $V^{(2)}(R,R';t)\equiv a_{R}^{\mu (t)} a_{R^{\prime
}}^{\mu (t)}$, to give two identical Crowd-Anticrowd `quasiparticles'
of charge-density $({{n_{R}}}-{{
{n_{\overline{R}}}}})$ which interact via a {\em time-independent} and {\em
position-independent} interaction term $V^{(2)}_{\rm eff}\equiv 1$.
The different types of Crowd-Anticrowd quasiparticle in Eq.
(\ref{final}) do not interact with each other, i.e. $({{n_{R}}}-{{
{n_{\overline{R}}}}})$ does not interact with $({{n_{R'}}}-{{
{n_{\overline{R'}}}}})$ if $R\neq R'$. Interestingly, this situation could
{\em not} arise in a conventional physical system containing collections of just two types of
charge (i.e. positive and negative).

A given numerical simulation will employ a given
strategy-allocation matrix (i.e. a given rank-$S$ tensor) $\Psi$.
As $R_{\rm max}$ increases from $1\rightarrow\infty$, $\Psi$ tends
to become increasingly disordered (i.e. increasingly non-uniform)
\cite{book,paul} since the ratio of the standard deviation to the
mean number of agents holding a particular set of $S$ strategies
is equal to $[({R_{\rm max}^S}-1)/N]^\frac{1}{2}$. There are two
regimes: (i) A `high-density' regime where $R_{\rm max}\ll N$.
Here the charge-densities 
$\{n_R\}$ tend to be large, non-zero values which monotonically
decrease with increasing $R$.  Hence the set $\{n_R\}$ acts like a
smooth function $n(R)\equiv \{n_R\}$.  (ii) A `low-density' regime
where $R_{\rm max}\gg N$. Here $\Psi$ becomes sparse with each
element $\Psi_{\rm R,R',R'',\dots}$ reduced to 0 or 1. The
$\{n_R\}$ should therefore be written as 1's or
0's in order to retain the discrete nature of the agents, and yet
also satisfy Eq. (\ref{vol}) \cite{book}. Depending on the
particular type of game, moving between regimes may or may not
produce an observable feature. In the MG, for example, $D_1$ does
not show an observable feature as $R_{\rm max}$ increases --
however $D_2$ does \cite{savit}. We leave aside the discussion as
to whether this constitutes a true phase-transition
\cite{others,paul} and instead discuss the explicit analytic
expressions for $D_2$ which result from Eq. (\ref{final}).  It is
easy to show that the mean number of agents using the $X$th most
popular strategy (i.e. after averaging over $\Psi$) is
\cite{book}:
\begin{equation} {{n_{X}}} = N \bigg[
\left( 1-\frac{(X-1)}{R_{\rm max}}\right) ^{S}-\left( 1-\frac{X}{R_{\rm max}}
\right) ^{S}\bigg] . \label{nav}
\end{equation}
The increasing non-uniformity in $\Psi$ as $R_{\rm max}$
increases, means that the popularity-ranking of $\overline R$
becomes increasingly independent of the popularity-ranking of $R$.
Using Eq. (\ref{nav}) with $S=2$, and averaging over all possible
$\overline R$ positions in Eq. (\ref{final}) to reflect the
independence of the popularity-rankings for $\overline R$ and $R$,
we obtain:
\begin{equation} D_2 ={\rm
Max}\bigg[\ \ {\frac{N^2}{3 R_{\rm max}}} \bigg(1-{R^{-2}_{\rm
max}}\bigg),\ \ N\bigg(1-\frac{N}{R_{\rm
max}}\bigg)\bigg]\label{lower}\ . \label{Dfin}
\end{equation}
The `Max' operation ensures that as $R_{\rm max}$ increases and
hence $\{n_R\}\rightarrow 0,1$, Eq. (\ref{vol}) is still satisfied
\cite{book}. Equation (\ref{Dfin}) underestimates $D_2$ at small
$R_{\rm max}$ (see Fig.~1) since it assumes that the rankings of
$\overline R$ and $R$ are unrelated, thereby overestimating the
Crowd-Anticrowd cancellation. By contrast, an overestimate of
$D_2$ at small $R_{\rm max}$ can be obtained by considering the
opposite limit whereby $\Psi$ is sufficiently uniform that the
popularity and strategy-performance rankings are identical. Hence
the strategy with popularity-ranking $X$ in Eq. (\ref{nav}) is
anticorrelated to the strategy with popularity-ranking $R_{\rm
max}+1-X$. This leads to a slightly modified first expression in
Eq. (\ref{Dfin}): $\frac{2 N^2}{3 R_{\rm max}} (1-R_{\rm
max}^{-2})$. Figure 1 shows that the resulting analytical
expressions reproduce the quantitative trends in the standard
deviation $D_{2}^{1/2}$ observed numerically for all $N$ and
$R_{\rm max}$, {\em and} they describe the wide spread in the numerical data
observed at small $R_{\rm max}$.

An important practical implication of the present paper is that the wide range of cluster-based approximation schemes developed in conventional many body-theory, might therefore usefully be extended to capture the dominant correlations in multi-agent competitive populations. Such generalizations will undoubtedly raise interesting issues for the Physics community, concerning the precise manner in which time-averages and configuration-averages should be performed within these traditional approximation schemes.


\begin{thebibliography}{0}

\bibitem{casti} J.L. Casti, {\em Would-be Worlds} (Wiley, New York, 1997). 

\bibitem{econo} A.C.C. Coolen, {\em The Mathematical Theory of Minority Games} (Oxford University Press, 2005); D. Challet, M. Marsili, Y.C. Zhang, {\em Minority Games} (Oxford University Press, 2004). See also http://www.unifr.ch/econophysics/minority.


\bibitem{rob} A. Soulier and T. Halpin-Healy, Phys. Rev. Lett. {\bf 90}, 258103
(2003); A. Bru, S. Albertos, J.A. Lopez Garcia-Asenjo, and I. Bru, Phys. Rev.
Lett. {\bf 92}, 238101 (2004); B. Huberman and R. Lukose, Science {\bf 277},
535 (1997); B. Arthur, Amer. Econ. Rev. {\bf 84}, 406 (1994); J.M. Epstein,
Proc. Natl. Acad. Sci. {\bf 99}, 7243 (2002). See also
the works of D. Wolpert and K. Tumer at http://ic.arc.nasa.gov

\bibitem{book} N.F. Johnson and P.M. Hui, cond-mat/0306516; N.F. Johnson, P.
Jefferies, P.M. Hui, {\em Financial Market Complexity} (Oxford University
Press, 2003).

\bibitem{others} D. Challet and Y.C. Zhang, Physica A {\bf 246}, 407 (1997); D.
Challet, M. Marsili and R. Zecchina, Phys. Rev. Lett. {\bf 82}, 2203 (1999);
J.A.F. Heimel, A.C.C. Coolen and D. Sherrington, Phys. Rev. E {\bf 65}, 016126
(2001); A. Cavagna, J.P. Garrahan, I. Giardina and D. Sherrington, Phys. Rev.
Lett. {\bf 83}, 4429 (1999).

\bibitem{savit} R. Savit, R. Manuca and R. Riolo, Phys. Rev. Lett. {\bf 82},
2203 (1999).

\bibitem{emg} N.F. Johnson, P.M. Hui, R. Jonson and T.S. Lo, Phys. Rev. Lett.
{\bf 82}, 3360 (1999); S. Hod
and E. Nakar, Phys. Rev. Lett. {\bf 88}, 238702 (2002); E. Burgos, H. Ceva, and
R.P.J. Perazzo, Phys. Rev. Lett. {\bf 91}, 189801 (2003); R. D'hulst and G.J.
Rodgers, Physica A {\bf 270}, 514 (1999).
These works focus on
probabilistic strategies.

\bibitem{us} N.F. Johnson, M. Hart and P.M. Hui, Physica A \textbf{269}, 1
(1999); M. Hart, P. Jefferies, N.F. Johnson and P.M.
Hui, Physica A \textbf{298}, 537 (2001); S.C. Choe, P.M. Hui and N.F.
Johnson, Phys. Rev. E {\bf 70}, 055101(R) (2004); M. Hart,
P. Jefferies, N.F. Johnson and P.M. Hui, Phys. Rev. E {\bf 63}, 017102 (2001).

\bibitem{paul} P. Jefferies, M.L. Hart and N.F. Johnson, Phys. Rev. E {\bf 65},
016105 (2002).

\bibitem{mahan} G.D. Mahan, {\em Many-Particle Physics}  (Plenum Publishing,
New York, 2000) 3rd Edition.

\bibitem{note} We interchange the order of the $\Psi$ and $t$-averaging
over the product of the $n_R$'s. Numerical simulations
suggest this is valid for the systems of interest.

\bibitem{lo} T.S. Lo, K.P. Chan, P.M. Hui, N.F. Johnson, Phys. Rev. E {\bf 71}, 050101(R) (2005).


\end{thebibliography}
\end{document}